\documentclass[preprint,aps,tightenlines,showpacs,nofootinbib]{revtex4}
\usepackage{latexsym}
\usepackage[dvips]{graphicx}
\usepackage[dvips]{color}
\newcommand{\beq}{\begin{eqnarray}}
\newcommand{\eeq}{\end{eqnarray}}
\newcommand{\bea}{\begin{eqnarray*}}
\newcommand{\eea}{\end{eqnarray*}}
\newcommand{\eq}{eqnarray}

\newcommand{\al}{{\alpha}}
\newcommand{\be}{{\beta}}
\newcommand{\ci}{\cite}

\newcommand{\ep}{{\epsilon}}

\newcommand{\la}{{\lambda}}
\newcommand{\La}{{\Lambda}}
\newcommand{\m}{{\mu}}

\newcommand{\si}{{\sigma}}

\newcommand{\ka}{{\kappa}}
\newcommand{\om}{{\omega}}

\newcommand{\pa}{{\partial}}
\newcommand{\no}{{\nonumber}}
\newcommand{\f}{\frac}
\newcommand{\ra}{\rightarrow}

\newcommand{\Sch}{Schwarzschild }

\newcommand{\fn}{\footnote}

\begin{document}

\preprint{arXiv:{1804.05698v3} [hep-th]}
\title{On Birkhoff's Theorem in
%and Scalar Gravitons
%in Ho\v{r}ava Gravity
%Lorentz Violating
%Renormalizable Quantum Gravity:\\
 %The %Four-Dimensional
 Ho\v{r}ava Gravity %Case
}

\author{Deniz O. Devecio\u{g}lu\footnote{E-mail address: dodeve@gmail.com}}
\author{Mu-In Park\footnote{E-mail address: muinpark@gmail.com, Corresponding author}}
\affiliation{ Research Institute for Basic Science, Sogang University,
Seoul, 121-742, Korea }
\date{\today}

\begin{abstract}
We study Birkhoff's theorem, which states the absence of time-dependent, spherically symmetric
vacuum solutions in four-dimensional Ho\v{r}ava gravity, which has been proposed as a renormalizable
quantum gravity without the ghost problem. We prove that the theorem is still valid 
for the usual
type of solutions %$(\dot{\be}=0)$
which admit the general relativity limit in the low energy (IR) region. However, for the unusual type
of solutions,
%$(\dot{\be}\neq 0)$,
it can be violated in high-energy (UV) region, due to the non-linear effects. %higher-spatial derivatives.
This implies that the scalar graviton can {\it emerge}
%exist
as the results of {\it non-linear} UV effects but is decoupled in IR regime.
%This provides a non-perturbative proof of {\it the non-linear %emergence of the scalar graviton in UV, but its absence or %decoupling in IR}.
An important implication of the non-linear, UV scalar graviton in Big Bang cosmology is also discussed.

\end{abstract}

\pacs{04.20.Cv, 04.30.-w, 04.60.-m, 04.20.Jb}

\maketitle

\newpage

%\section{Introduction}

Birkhoff's theorem, which states the absence of time-dependent, spherically symmetric vacuum solutions, is an important consequence of general relativity (GR) \ci{Jebs,Birk}. This implies that the uniqueness of a spherically symmetric solution as the static one, given by the \Sch solution. Furthermore, it also implies the absence of gravitational radiation for pulsating or collapsing, spherically symmetric bodies, which can be stated as the absence of spin-0, or {\it scalar gravitons} in modern terms. The Birkhoff's theorem is also quite important for the study of cosmological as well as astrophysical problems \ci{Wein}.

On the other hand, it is also well known that GR would not be a UV complete theory due to
lack of renormalizability. Several years ago, Ho\v{r}ava proposed a renormalizable,
higher-derivative gravity theory, without the ghost problem in the usual covariant
higher-curvature gravities,
%which reduces to Einstein gravity in IR but with improved UV behaviors,
%by abandoning Einstein's equal-footing treatment of space and time through
by considering different scaling dimensions for space and time
%, $[t]=-1, [{\bf x}]=-z$ with the dynamical critical exponents $(z>1)$
\ci{Hora}. So, it would be an important question whether the Birkhoff's theorem can be still valid or
needs to be modified for this quantum gravity model. Actually, since there have been long standing
debates about the scalar graviton mode and the recovery of GR in IR, the study of Birkhoff's theorem is
also directly related to the fundamental question about the consistency of Ho\v{r}ava
gravity \ci{Char,Blas:0906,%Koya,
Blas:0909%,Papa
}.

To this ends, we start by considering the ADM decomposition of the
metric
\begin{\eq}
ds^2=-N^2 dt^2+g_{ij}\left(dx^i+N^i dt\right)\left(dx^j+N^j
dt\right)\
\end{\eq}
and the Ho\v{r}ava gravity action\fn{For $\la = 1/3$,
%the action is not defined and
a separate consideration with different coefficients is needed \ci{Park:0910b}. We will briefly mention about the results for this case later.}, which is power-counting renormalizable in four dimensions \ci{Hora},
\begin{\eq}
\label{HL action}
S &=& \int d \eta d^3x \sqrt{g} N \left\{ \frac{2}{\kappa^2}
\left(K_{ij}K^{ij} - \lambda K^2 \right) - {\cal V}[g_{ij}] \right\},\\
-{\cal V}[g_{ij}]&=&\frac{\kappa^2\mu^2[(\Lambda_W-\om) R -3\Lambda_W^2]}{8(1-3\lambda)}+
\frac{\kappa^2\mu^2 (1-4\lambda)}{32(1-3\lambda)}R^2
-\frac{\kappa^2}{2\nu^4} \left(C_{ij} -\frac{\mu \nu^2}{2}R_{ij}\right)
\left(C^{ij} -\frac{\mu \nu^2}{2}R^{ij}\right) \no
, \label{V_Horava}
\end{\eq}
where
$
%\begin{\eq}
 K_{ij}=%\frac{1}{2N}
(2N)^{-1} \left(\dot{g}_{ij}-\nabla_i N_j-\nabla_jN_i\right)
% \end{\eq}
$
is the extrinsic curvature,
$
%\begin{\eq}
 C^{ij}=\epsilon^{ik\ell}\nabla_k
%\left
(R^{j}{}_\ell-%\frac{1}{4}
R^{} \delta^j_\ell/4
%\right
)\
 %\end{\eq}
$
is the Cotton tensor, $\kappa,\lambda,\nu,\mu, \La_W$
are coupling parameters, and $\om$ an IR-modification parameter which breaks softly the detailed
balance condition in IR \ci{Hora,Keha,Park:0905%,Nast
} so that Newton's gravity or GR limit exits, without changing the improved UV behaviors.
[ $\epsilon^{ik\ell}$ is the Levi-Civita symbol, $R_{ij}$ and $R$ are the three-dimensional (Euclidean) Ricci tensor and scalar, respectively ]

In addition to these standard terms of the action, for completeness, we will also consider the
extension terms which depend on the proper acceleration $a_i=\partial_i ln N$  \ci{Blas:0909}, which
have been introduced to avoid the problems raised in \ci{Char,Blas:0906%,Koya
}.
% of the standard action (\ref{HL action}).
Since the key role of the extension terms is in IR, we will consider the modification of ${\cal V}[g_{ij}]$ by \begin{\eq}
{\cal \delta V}[g_{ij},a_i]=-\f{\si}{2} a_i a^i, \label{ai_extension}
\end{\eq}
which is the only relevant term in IR \ci{Blas:0909}.

In order to study Birkhoff's theorem, let us consider the spherically symmetric, time dependent metric with the ansatz,
\begin{\eq}
ds^{2}=-e^{2 \alpha (t,r)}dt^2+e^{2\beta(t,r)}dr^{2}+r^2 (d\theta^2 +\mbox{sin}^2\theta d \phi^2).
\label{metric_ansatz}
\end{\eq}
Then, the Hamiltonian and momentum constraints are reduced to
\beq
{\cal H}&=&-\f{2 (1-\la)}{\kappa^2} \dot{\be}^2 e^{-2 \alpha} +\frac{\kappa^2\mu^2}{8(1-3\lambda)} \left[\f{2(\Lambda_W - \omega)e^{-2\be} }{r^2} \left(2 r \be'+e^{2\be}-1 \right)  -3\Lambda_W^2 \right] \no \\
&&+\frac{\kappa^2\mu^2 e^{-4\be}}{8(1-3\lambda) r^4}\left\{ -2(1-\la) r^2 \be'^2-4 \la r \be' (e^{2\be}-1)-(1-2 \la) (e^{2\be}-1)^2\right\} \no \\
&&-\si e^{-2\be} \left(\f{1}{2} \al'^2 +\al'' +\f{2}{r} \al' -\al' \be' \right) =0\,
,\label{eom1_ansatz} \\
{\cal H}^r &=&\f{e^{-(\al+2\be)}}{r} \left[ \left(2-r(1-\la) \al'\right) \dot{\be}+r (1-\la) {\dot\be}'\right]=0\, ,\label{eom2_ansatz}
\eeq
and the field equations of $g^{ij}$ are reduced to a single equation (see Appendix {\bf A} for the details)
\beq
E_\be
%&\equiv& e^{-\al} E_{11} +\f{1}{2}e^{2\be} {\cal H} \label{eom3_ansatz} \\
 &=&\f{2 (1-\la)}{\kappa^2} (\ddot{\be}-\dot{\al}\dot{\be}) e^{2 (\be-\al)}
 +\frac{\kappa^2\mu^2 (\Lambda_W - \omega)}{4(1-3\lambda)r} \left(\al'+\be'\right)
 -\f{\si}{2} \left(\al'' +\f{2}{r} \al' -\al' \be' \right)
 \no \\
 &&+\frac{\kappa^2\mu^2 e^{-2\be}}{4(1-3\lambda) r^4}
 \left\{ (1-\la)\left[ r^2 (\be'^2-\be''-\al'\be')+(e^{2\be}-1)\right]
 -\la r (\al'+\be')(e^{2\be}-1)\right\} =0\, .
 \label{eom3_ansatz}
 \eeq

In GR case, where $\la=1, \si=0$, and the higher-derivative terms ($\{ \cdots\}$ terms
in (\ref{eom1_ansatz}) and (\ref{eom3_ansatz})) are absent, it is easy to see that one can
obtain the unique solution $\dot{\be}=0$ from ${\cal H}_i=0$,
%of (\ref{eom2_ansatz})
and $\dot{\al}'=0$ from $\dot{\cal H}=0$, which tells the {\it time-independence of metric} (\ref{metric_ansatz}) and so proves the Birkhoff's theorem: $\dot{\al}'=0$  implies $\al(t,r)=a(t)+b(r)$ but $a(t)$ can be removed by redefining the time coordinate $t$.

Generally, however, there could exist solutions which may break the theorem due to, either  the IR Lorentz violation from $\la \neq 1$ or $\si \neq 0$, or the UV Lorentz violation from higher- derivative terms. We accordingly classify the solutions largely by the time dependency of $\be$, {\it i.e.}, $\dot{\be}=0$ or $\dot{\be}\neq0$.

\subsection{Case $\dot{\be}=0$:}

For  $\la=1$, this is the only possible solution of the momentum constraint (\ref{eom2_ansatz}), as in GR case, though not the unique solution for $\la \neq 1$. However, either $\la=1$ or $\la \neq 1$, one can prove $\dot{\al}'=0$, {\it i.e.}, admitting Birkhoff's theorem, for generic values of coupling parameters, except one particular set of parameters which relates UV and IR. To prove this, we consider the time derivatives of (\ref{eom1_ansatz}) and (\ref{eom3_ansatz}), which reduce to

\beq
\dot{\cal H}&=&
-\si e^{-2\be}
\left[
\left(\al'-\be' +\f{2}{r}\right) \dot{\al}' +\dot{\al}''
\right] =0\, ,\label{eom1_dot} \\
\dot{E}_\be
&=&\frac{\kappa^2\mu^2 e^{-2\be}}{4(1-3\lambda) r^4}
 \left[ -\la r(e^{2\be}-1)-(1-\la) r^2 \be' +(\La_W-\om) r^3 e^{2\be}\right] \dot{\al}'  -\f{\si}{2}
\left[
\left(-\be' +\f{2}{r}\right) \dot{\al}' +\dot{\al}''
\right] \no \\
 &=& \left\{\f{\si}{2} \al'+\frac{\kappa^2\mu^2 }{4(1-3\lambda) r^4}
 \left[ -\la r(1-e^{-2\be})-(1-\la) r^2 e^{-2\be}\be' +(\La_W-\om) r^3 \right]  \right\} \dot{\al}' =0\,
 ,\label{eom3_dot}
 \eeq
where we have used (\ref{eom1_dot}) in the last step of (\ref{eom3_dot}).

In the presence of the extension term (\ref{ai_extension}), {\it i.e.}, $\si \neq 0$, one finds that $\dot{\al}'=0$ is the only possible solution so that Birkhoff's theorem is satisfied: For the case of $\{ \cdots \}=0$ in (\ref{eom3_dot}), $\al(t,r)$ can be integrated as $\al(t,r)=a(t)+b(r)$ so that $\dot{\al}'=0$ is satisfied again.

On the other hand, in the absence of the extension, {\it i.e.}, $\si=0$, (\ref{eom3_dot}) gives the usual solution $\dot{\al}'=0$, or an unusual solution for $\be(t,r)$,
\beq
\be(t,r)=-ln\sqrt{1+(\om-\La_W)r^2 +C r^{\f{2 \la}{\la-1}} },
\label{LMP_type sol}
\eeq
which makes $\{ \cdots \}=0$ in (\ref{eom3_dot}), {\it even without knowing $\dot{\al}'$}, where $C$ is an integration constant, which corresponds to the mass for the static case. In general, the second, unusual solution (\ref{LMP_type sol}) is not compatible with the Hamiltonian constraint (\ref{eom1_ansatz}), which reduces to

\beq
{\cal H}=\frac{\ka^2\m^2 }{8(1-3\la)(1-\la) r^4}
\left[ -(1-3 \la) C^2 r^{\f{4 \la}{\la-1}}+3 \om (\om-2 \La_W )(1-\la)r^4\right]=0.
\eeq
However, there exists one exceptional, compatible solution when the conditions
\beq
C=0,~  \om (\om-2 \La_W )=0
\label{LMP_type}
\eeq
are satisfied.\footnote{Note that, with the conditions in (\ref{LMP_type}), the solution (\ref{LMP_type sol}) is valid for arbitrary $\la$, including the $\la=1$ case: Actually, with $C=0$, the solution corresponds to the {\it zero-mass} limit of the $\la=1$ static black hole solution in \ci{Park:0905}.}

There are two possible solutions for the second condition in (\ref{LMP_type}), {\it i.e.}, $\om=0$
or $\om=2 \La_W$. The first case, $\om=0$ corresponds to the solution without the
IR-modification
term %that has been first studied in
\cite{Lu}. The second case, $\om=2 \La_W$ is the  corresponding {\it new} solution but, at this time, with the IR modification. A curious property of these solutions is that $\al(t,r)$ is not constrained by the equations of motions so that {\it  $\al(t,r)$ can be an arbitrary function of space and time.} It is this later property that allows the time dependence in the metric and so {\it could} violate the Birkhoff's theorem, even though $\dot{\be}=0$ from (\ref{LMP_type sol}) and (\ref{LMP_type}). However, we note that these solutions do not have the GR limit, $\la \ra 1, ~\m \ra 0,~ \om \ra \infty,~ \La_W \ra \infty$ with ` $\m^2 \om,~ \m^2 \La_W \sim \mbox{fixed}$ ' \ci{Park:1508} so that Birkhoff's theorem could be violated but only in the non-GR branch.\footnote{The solution for $\om=0$ seems to be due to the UV detailed balance condition since it does not exist for more generic UV actions \ci{Lu}. However, the solution for $\om=2 \La_W$ seems to be more generic since one can also obtain the similar solution with modified choice of $\om=\om(\La_W)$ for more generic UV terms.} Another peculiar property of the solution for $\om=\om(\La_W)$ is that it requires some correlations between IR and UV terms so that only their combined equations have the solution, though separate ones do not.

\subsection{Case $\dot{\be}\neq0$:}

For $\la \neq 1$, in addition to the usual solution $\dot{\be}=0$, one may also consider the case
of $\dot{\be}\neq 0$ generally, which breaks the  Birkhoff's theorem manifestly from the momentum
constraint (\ref{eom2_ansatz}). In order to see whether this is really a
%possible
feasible solution or not, we need to see whether it is compatible with the Hamiltonian constraints and other field equations. Actually, in this case $\al (t,r)$ can be determined as
\beq
\al'(t,r)=\f{2}{r (1-\la)} +\f{\dot{\be}'}{\dot{\be}}
\label{alpha_eq}
\eeq
from (\ref{eom2_ansatz}), and be integrated
%to give the general solution
as
\beq
\al(t,r)= ln \left( \dot{\be}(t,r) r^{\f{2}{1-\la}}\right) +a(t),
\label{alpha_sol}
\eeq
with one undetermined, time-dependent function $a(t)$. Then, after some computations with this general solution (\ref{alpha_sol}), one can find that Hamiltonian constraint (\ref{eom1_ansatz}) reduces, for the standard case of $\si=0$,
\beq
&&\f{16 (1-\la)(1-3 \la)}{\ka^4 \m^2}r^{\f{4}{\la-1}} e^{-2 a(t)}=
\f{2(\La_W -\om)}{r^2}\left(-r f'+1-f \right)-3 \La_W^2  \no \\
&&~~~~~~~~~~~~~~~~~+\f{1}{r^4} \left\{-\f{1}{2}(1-\la)r^2 f'^2 +2 \la r f' (1-f)-(1-2 \la) (1-f)^2 \label{eom2_ansatz_second case}
\right\},
\eeq
where we have introduced $f(t,r) \equiv
e^{-2 \be(t,r)}$ for convenience. {Then, (\ref{eom2_ansatz_second case}) and (\ref{eom3_ansatz}) become the equations for $\beta(t,r)$ and $a(t)$.} Once we get the solutions for (\ref{eom2_ansatz_second case}), one
can prove or disprove the Birkhoff's theorem by checking whether it can still satisfy the remaining
field equation (\ref{eom3_ansatz}),
 %, which reduces to (\ref{})
or not. To this ends, instead of getting the {\it explicit} solutions for the full equations with both UV and IR parts, which is a formidable task for our case of $\la \neq 1$, we consider the limiting solutions for the UV and IR equations separately.

First, for the Hamiltonian equation (\ref{eom2_ansatz_second case}) with only the UV terms ($\{\cdots \}$
terms in the right hand side), which corresponds to the limiting case of $\om, \La_W \ra 0$ of the full
equation (\ref{eom2_ansatz_second case}), one can solve the equation explicitly and finds that there are
two general solutions but only one solution is compatible with the field equation (\ref{eom3_ansatz}),
which is given by\footnote{In IR region of large $r$, %$r \gg 1$,
the solution approaches to {
$f=e^{-2 \be}=1$ for $\la<1$,}
%,~N^2=e^{2 \al}=0$,
which is consistent with the case of $\la=1$ in Sec. {\bf A}. In other words, there is no discontinuity at
{the limit $\la \ra 1_{-}$} in IR,
%as in
{which may be comparable to} the {\it Vainshtein mechanism} for the {\it massless} limit of massive gravity
\ci{Vain}.}
,
\beq
f(t,r) =1 \pm \f{4|\la-1|r^{\f{2 \la}{\la-1}}}{\ep \sqrt{-\ka^4 \m^2}}e^{-a(t)},
\label{UV__time_sol_1}
\eeq
where $\ep \equiv \mbox{sign} (3 \la-1)$.
This solution is peculiar in that {$a(t)$ is unconstrained and} there is no integration constant, and this is due to a factorization of an algebraic equation, called Abel's equation, in the UV limit of $\om, \La_W \ra 0$ \ci{Kamk}. {However,} from our starting assumption, we find $\dot{\be} \propto \dot{a} \neq 0$, {\it i.e.}, $a(t)$ can not be trivially a constant so that the solution manifestly violates the Birkhoff's theorem in the UV regime.\fn{For the case of $\la=1/3$, where a separate analysis is needed, we find that there is no corresponding time-dependent solution in UV \ci{Prog}. This is the only result which is qualitatively different from the case of $\la \neq 1/3$ and it would be probably due to the additional (anisotropic) Weyl symmetry in UV \ci{Hora,Park:0910b}.}

On the other hand, from the relation (\ref{alpha_eq}) or (\ref{alpha_sol}), $\al(t,r)$ is given by
\beq
\al(t,r)=ln \left[ \f{2|\la-1| r^2 \dot{a}(t)}{\pm \ep \sqrt{-\ka^4 \m^2}+4 |\la-1| r^{\f{2\la}{\la-1}}e^{-a(t)} }\right].
\label{UV__time_sol_2}
\eeq
Moreover, it is interesting to note that the solutions (\ref{UV__time_sol_1}) and (\ref{UV__time_sol_2}) exist only for the {\it de-Sitter} branch with a positive cosmological constant $\La \propto -\m^2 >0$, like our current accelerating Universe \ci{Ries:1998}.

Second, for the Hamiltonian equation (\ref{eom2_ansatz_second case}) with only the IR terms (the first two terms in the right hand side), one can also get the explicit solution but find that it is not compatible with the field equation (\ref{eom3_ansatz}) either. So, even for the $\la \neq 1$ case with an IR Lorentz violation, there is no time-dependent, {\it i.e.}, Birkhoff's theorem violating, solution in IR regime.

Finally, with the %$a_i$
extension terms in (\ref{ai_extension}), the Hamiltonian equation (\ref{eom2_ansatz_second case}) has additional contributions
\beq
-\f{8 (1-3 \la) \si}{\ka^2 \m^2}f \left[U'+\f{1}{2}(U^2+U J)
+\f{2}{r}\left(\f{2-\la}{1-\la}\right)U
+\f{J}{r(1-\la)}
+\f{2}{r^{2}}\f{(2-\la)}{(1-\la)^2}\right],
\label{eom2_ansatz_second case_ai}
\eeq
where we have introduced $U(t,r) \equiv (ln \dot{\be})',~
%[ln (\dot{f}/f )]',
J(t,r) \equiv -2 \be'$ so that the original PDE problem is reduced to an ODE problem with respect to $r$, with a fixed time $t$.
%f'/f.$
With the extension terms, solving even the IR equation of (\ref{eom2_ansatz_second case}), which actually shows the key role of %$a_i$
the extension terms, is a difficult task and its general solution is not available. However, there exists a simple situation that allows an exact solution, which is now compatible with the equation (\ref{eom3_ansatz}) [we will omit the detailed derivation, which is quite cumbersome],
\beq
\al(t,r)=ln \left(\f{8\dot{C}_{1}(t)}{r \sqrt{-\ka^4 \m^2 \om}}\right), ~~
\be(t,r)=\f{C_1(t)}{r^2}+b(r),\no
\eeq
with $\la=-1, \La_W=0, \si=-\m^2 \ka^2 \om/8, a(t)=-ln\sqrt{-\ka^4 \m^2 \om/64}$ and two arbitrary functions $C_1(t)$ and $b(r)$. From our stating assumption of $\dot{\be} \neq0$, $C_1(t)$ can not be just a constant so that this solution violates the Birkhoff's theorem manifestly even in IR. This situation is quite different from the case of $\dot{\be}=0$ and the other cases of $\dot{\be}
\neq 0$ with $\si=0$, where the violations of Birkhoff's theorem may occur as the
UV effects (or combined UV/IR effects for the case of $\dot{\be}=0$). For other
values of parameters, we have obtained the (time-dependent) solutions
{\it numerically},
by solving
%(\ref{eom3_ansatz}) and (\ref{eom2_ansatz_second case}) with
%(\ref{eom2_ansatz_second case_ai})
about $\be(t,r)$ and $\dot{\be}(t,r)$ using {\it Mathematica}. (Fig.1)
%and the results seem to show some correlations between the parameters
%(especially, $\la$ and $\om$) for the existence of solutions.

\begin{figure}
\includegraphics[width=9cm,keepaspectratio]{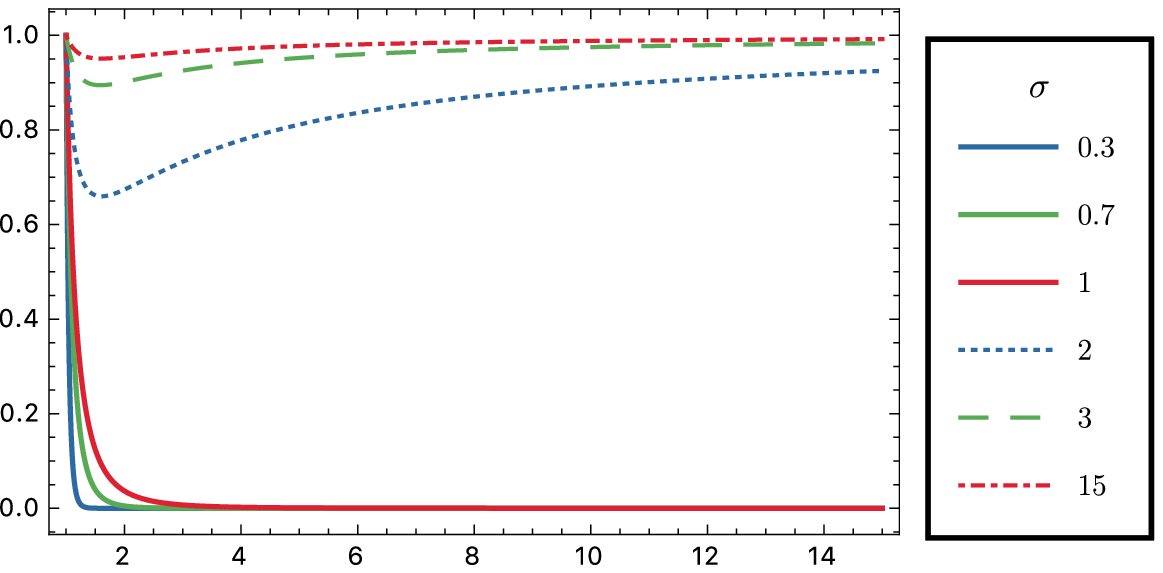}
\includegraphics[width=7cm,keepaspectratio]{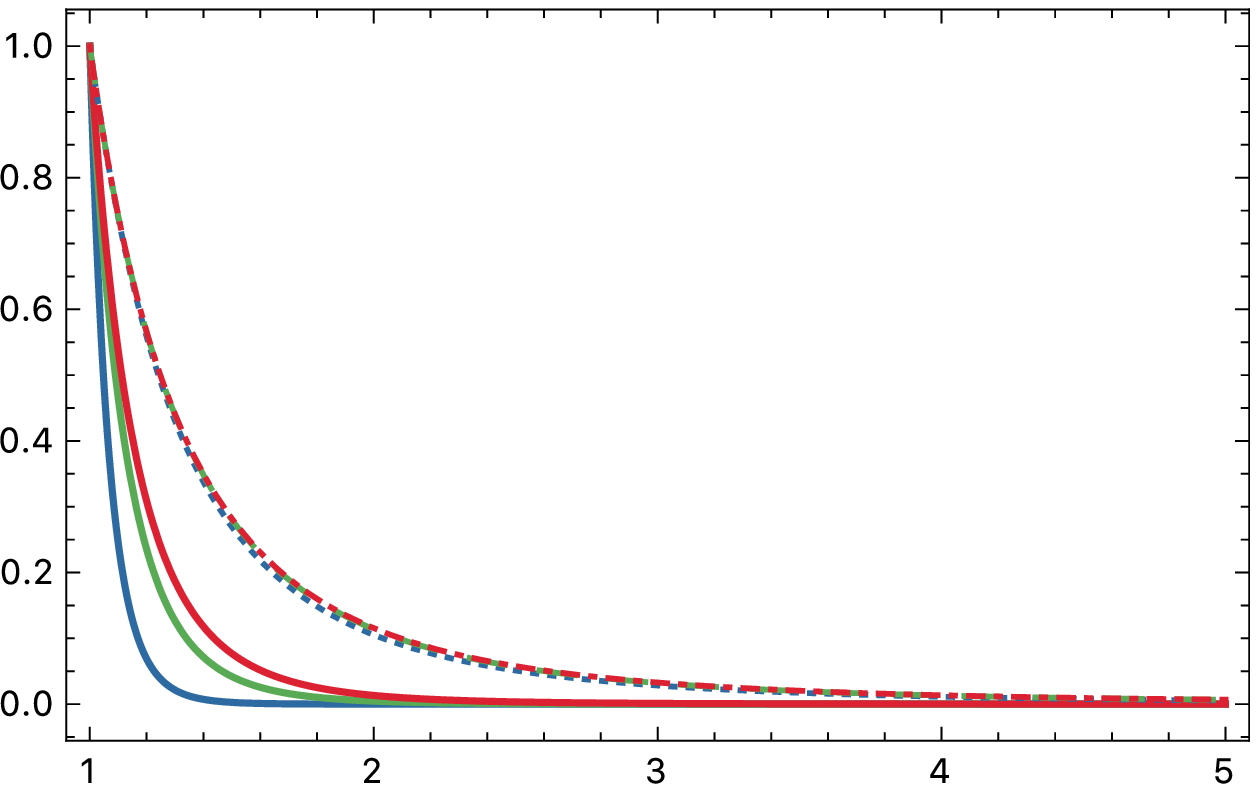}
\caption{Plots of numerical solutions for $f(t,r)^{-1}=e^{2 \be(t,r)}$
(left), $\dot{\be}(t,r)$
%\dot{\be}'/\dot{\be}$
(right) vs.
$r$ for varying $\si$, at $t=t_0$. %=\mbox{fixed}$. }
Here, we have considered
$\la=0.35,~\Lambda_W=0,~\om=0.225,~\mu=3,~\ka=1$,
~$a(t_0)=e^3, \dot{a}(t_0)=0$. These show two different branches of
solutions with different asymptotes, $f=1$ for $\si >1$ (upper curves) or
$f=\infty$ %depending on
for $\si \leq 1$ (lower curves).}
%The upper and lower curves correspond to non-tachyonic and tachyonic ghosts, respectively, in the linear perturbations \ci{Blas:0909}. }
\label{fig:Temp_AdS}
\end{figure}

In conclusion, we have proved that, for the standard form of Ho\v{r}ava gravity, Birkhoff's theorem is
satisfied in IR {for the usual
type of solutions %$(\dot{\be}=0)$
which admit the GR limit} but can be violated in UV { for the unusual
type of solutions}. %due to {\it non-linear} effects.
In relation to the gravitational radiations, this implies that the scalar gravitons {could}
exist as the results of
%non-linear
UV effects but {be} decoupled in IR regime.
Here, it is important to note that we have considered the problems with the full
non-linearity by obtaining the exact (time-dependent) solutions for the
on-linear equations of motion. This result is consistent with the (fully
non-linear) constraint analysis \ci{Bell:1004}, which has been thought to
be inconsistent \ci{Li,Henn}, but in contrast to the absence of the scalar
gravitons for the whole UV and IR energy ranges in the linear perturbation analyses
%In other words, our result corresponds a `` non-perturbative" proof
%of the results in
\ci{Park:0910,Kim,Gao,Gong,Shin:1701}. This implies a remarkable fact that {\it the UV emergence of time-dependent solutions, {\it i.e.,} violation of Birkhoff's theorem, or equivalently scalar gravitons, are the non-linear effect in UV.} Actually, if we consider small $a(t)$, the UV time-dependent solution (\ref{UV__time_sol_1}), (\ref{UV__time_sol_2}) can be expanded as
\beq
f^{-1}&=&e^{2 \be}=(1+\zeta^{-1})(1+ {a(t)}{\zeta}^{-1}+
{ {\cal O} (a(t)^2 )}
%\cdots
), \label{UV__time_sol_1_expand}\\
N^2&=&e^{2 \al}= \f{1}{4}{r^{\f{4}{1-\la}} \dot{a}^2(t)}{ \zeta^{-2}}[1 +{2a(t)}{\zeta}^{-1}
+{ {\cal O} (a(t)^2)}
%\cdots
], \label{UV__time_sol_2_expand}
\eeq
where $\zeta=1 \pm \f{\ep \sqrt{-\ka^4 \m^2}}{4 |\la-1|}r^{\f{-2 \la}{\la-1}}$.
This shows explicitly that, {when we consider small $\zeta^{-1}$ as well}, $a(t)$ does not appear at the leading, linear orders but emerges only at the sub-leading, {\it i.e.}, non-linear orders, in consistently with
the constraint analysis \ci{Bell:1004}. [Note that, at the {leading}
%linear
order, the $\dot{a}^2(t)$ factor in (\ref{UV__time_sol_2_expand}) can be removed by redefining the time as $dt \ra dt'=dt/\dot{a}(t)$.]

On the other hand, for the extended Ho\v{r}ava gravity with %$a_i$
the term of (\ref{ai_extension}), the Birkhoff's theorem is still satisfied for $\dot{\be}=0$, {\it i.e.}, %one can get
$\dot{\al}'=0$ so that there is no time-dependent solutions for the full theory with both the UV and IR terms. However, we have shown an explicit time-dependent solution which violates the theorem in IR for $\dot{\be} \neq 0,~ \la \neq 1$. This is consistent with the perturbative \ci{Blas:0909} as well as non-perturbative %(constraint)
analyses \ci{Bell:1106} but this seems to be {\it potentially} problematic since it implies the existence of gravitational radiations even for pulsating or collapsing, spherically symmetric bodies in IR, which have not been detected yet; even more, it  does not reproduce the GR or Newton's gravity limit in IR, which has been well tested \ci{Ryes:1003}.

On the contrary, the existence of a scalar graviton mode, which seems to be represented by {a scalar} function $a(t)$, { indicating the {\it instantaneous} mode}, in the general solution (\ref{alpha_sol}), would have an important role in cosmology. Usually, we need (at least one) primordial scalar matter field, {with an {\it instantaneous} background,}  in order to accommodate the observed (nearly scale invariant) scalar power spectrum in CMB data within the inflationary theory \ci{Guth}. But, now the scalar degree of freedom which is inherent in the non-linear UV regime of Ho\v{r}ava gravity could have a similar role of the primordial scalar field in the early Universe. It would be an outstanding question whether Ho\v{r}ava gravity can provide a consistent framework for the Big Bang cosmology without introducing the artificial primordial scalar field and inflationary scenario \ci{Shin:1701}.

\section*{Acknowledgments}
We would like to thank G{\"o}khan Alka\c{c} for helpful discussions on numerical analysis.
This work was supported by Basic Science Research Program through the National Research Foundation of Korea (NRF) funded by the Ministry of Education, Science and Technology {(2016R1A2B401304)}.

\appendix

\begin{section}
{Full equations of motion}
\end{section}
The Hamiltonian and momentum constraints, following from the variations of the action (\ref{HL action}) with the extension term (\ref{ai_extension}) for $N$ and $N^i$ respectively, are given by
\beq
{\cal H}&\equiv &-\f{2}{\kappa^2}(K_{ij}K^{ij} -\lambda K^2) +\frac{\kappa^2\mu^2 \left[(\Lambda_W - \omega) R -3\Lambda_W^2 \right]}{8(1-3\lambda)}
+\frac{\kappa^2\mu^2 (1-4\lambda)}{32(1-3\lambda)}R^2 -
\frac{\kappa^2}{2\nu^4} Z_{ij} Z^{ij}\no \\
&&+{\si}\left(\f{1}{2}\f{\nabla_i N \nabla^i N}{N^2}-\f{\nabla_k \nabla^k N}{N}\right)=0\, ,\label{eom1} \\
{\cal H}^i &\equiv&\nabla_k(K^{ki}-\lambda\,Kg^{ki})=0\, ,\label{eom2}
\eeq
where
\begin{\eq}
Z_{ij}\equiv C_{ij} - \frac{\mu \nu^2}{2} R_{ij}.
\end{\eq}

The equations of motion from variation of $\delta g^{ij}$ are given by \ci{Lu,Kiri,Alie%,Lee:1110
}
\beq
E_{ij}&\equiv& \frac{2}{\kappa^2}E_{ij}^{(1)}-\frac{2\lambda}{\kappa^2}E_{ij}^{(2)}
+\frac{\kappa^2\mu^2(\Lambda_W-\omega)}{8(1-3\lambda)}E_{ij}^{(3)} +\frac{\kappa^2\mu^2(1-4\lambda)}{32(1-3\lambda)}E_{ij}^{(4)}-\frac{\mu\kappa^2}{4\nu^2}E_{ij}^{(5)}
-\frac{\kappa^2}{2\nu^4}E_{ij}^{(6)}\no \\
&&+\frac{\si}{2}E_{ij}^{(7)}=0, \label{eom3}
\eeq
where
\bea
E_{ij}^{(1)}&=& N_i \nabla_k K^k{}_j + N_j\nabla_k K^k{}_i -K^k{}_i
\nabla_j N_k-
   K^k{}_j\nabla_i N_k - N^k\nabla_k K_{ij}\no\\
&& - 2N K_{ik} K_j{}^k
  -\frac{1}{2} N K^{k\ell} K_{k\ell}\, g_{ij} + N K K_{ij} + \dot K_{ij}
\,,\no \\
E_{ij}^{(2)}&=& \frac{1}{2} NK^2 g_{ij}+ N_i \pa_j K+
N_j \pa_i K- N^k (\pa_k K)g_{ij}+  \dot K\, g_{ij}\,,\no\\
E_{ij}^{(3)}&=&N\left(R_{ij}- \frac{1}{2}R g_{ij}+\frac{3}{2}
\frac{\Lambda_W^2}{\Lambda_W-\omega} g_{ij}\right)-(
\nabla_i\nabla_j-g_{ij}\nabla_k\nabla^k)N\,,\no\\
E_{ij}^{(4)}&=&NR\left(2 R_{ij}-\frac{1}{2}R g_{ij}\right)- 2
\big(\nabla_i\nabla_j-g_{ij}\nabla_k\nabla^k\big)(NR)\,,\no\\
E_{ij}^{(5)}&=&\nabla_k\big[\nabla_j(N Z^k_{~~i}) +\nabla_i(N
Z^k_{~~j})\big]  -\nabla_k\nabla^k(N Z_{ij})
-\nabla_k\nabla_\ell(N Z^{k\ell})g_{ij}\,, \no \\
%\eea
%\bea \label{eeeq}
E_{ij}^{(6)}&=&-\frac{1}{2}N Z_{k\ell}Z^{k\ell}g_{ij}+
2NZ_{ik}Z_j^{~k}-N(Z_{ik}C_j^{~k}+Z_{jk}C_i^{~k})
+NZ_{k\ell}C^{k\ell}g_{ij}\no\\
&&-\frac{1}{2}\nabla_k\big[N\epsilon^{mk\ell}
(Z_{mi}R_{j\ell}+Z_{mj}R_{i\ell})\big]
+\frac{1}{2}R^n{}_\ell\, \nabla_n\big[N\epsilon^{mk\ell}(Z_{mi}g_{kj}
+Z_{mj}g_{ki})\big]\no\\
&&-\frac{1}{2}\nabla_n\big[NZ_m^{~n}\epsilon^{mk\ell}
(g_{ki}R_{j\ell}+g_{kj}R_{i\ell})\big]
-\frac{1}{2}\nabla_n\nabla^n\nabla_k\big[N\epsilon^{mk\ell}
(Z_{mi}g_{j\ell}+Z_{mj}g_{i\ell})\big]\no\\
&&+\frac{1}{2}\nabla_n\big[\nabla_i\nabla_k(NZ_m^{~n}\epsilon^{mk\ell})
g_{j\ell}+\nabla_j\nabla_k(NZ_m^{~n}\epsilon^{mk\ell})
g_{i\ell}\big]\no\\
&&+\frac{1}{2}\nabla_\ell\big[\nabla_i\nabla_k(NZ_{mj}
\epsilon^{mk\ell})+\nabla_j\nabla_k(NZ_{mi}
\epsilon^{mk\ell})\big]
-\nabla_n\nabla_\ell\nabla_k
(NZ_m^{~n}\epsilon^{mk\ell})g_{ij}\,, \\
E_{ij}^{(7)}&=&\f{1}{N} \left( -\f{1}{2} g_{ij} \nabla_i N \nabla^i N +\nabla_i N \nabla_j N\right).
\eea

In general, from the spherical symmetry, there are %only
two non-vanishing field equations $E_{rr}, E_{\theta \theta}=E_{\phi \phi}/(r^2 sin \theta)$ but one finds that there is only one independent equation due to a (remarkable) relation $
2 e^{\be} E_{\theta \theta}-2 r^2 e^{-\be} E_{rr} -2 (e^{-\be})' r^3 E_{rr} -r^3 e^{-\be} (E_{rr})'+2 r^3 \f{\pa}{\pa t}(e^{3 \be} {\cal H}^r)/\ka^2-r^3 e^{\be} (e^{\al})' {\cal H}/2=0.$

 And $E_\be$ in (\ref{eom3_ansatz}) is given by
\beq
E_\be= e^{-\al} E_{rr} +\f{1}{2}e^{2\be} {\cal H}. \label{eom3_ansatz_Appendix}
\eeq

%%%%%%%%%% References %%%%%%%%%%%%%%%%%%%%%%%%%
\newcommand{\J}[4]{#1 {\bf #2} #3 (#4)}
\newcommand{\andJ}[3]{{\bf #1} (#2) #3}
\newcommand{\AP}{Ann. Phys. (N.Y.)}
\newcommand{\MPL}{Mod. Phys. Lett.}
\newcommand{\NP}{Nucl. Phys.}
\newcommand{\PL}{Phys. Lett.}
\newcommand{\PR}{Phys. Rev. D}
\newcommand{\PRL}{Phys. Rev. Lett.}
\newcommand{\PTP}{Prog. Theor. Phys.}
\newcommand{\hep}[1]{ hep-th/{#1}}
\newcommand{\hepp}[1]{ hep-ph/{#1}}
\newcommand{\hepg}[1]{ gr-qc/{#1}}
\newcommand{\bi}{ \bibitem}
%%%%%%%%%%%%%%%%%%%%%%%%%%%%%%%%%%%%%%%%%%%%%%%

\end{document}